\documentclass[aps,preprint]{revtex4-1}
\usepackage[utf8]{inputenc}
\usepackage[english]{babel}
\usepackage{amsmath}
\usepackage{amssymb}
\usepackage{float}
\usepackage{graphicx}
\usepackage{hyperref}

\begin{document}

\title{\textbf{Combined electronic excitation and knock-on damage in monolayer MoS$_2$}}
\author{Carsten Speckmann$^{1,2,*}$, Julia Lang$^{1}$, Jacob Madsen$^1$,\\ Mohammad Reza Ahmadpour Monazam$^1$, Georg Zagler$^{1,2}$, Gregor T. Leuthner$^{1,2}$,\\ Niall McEvoy$^{3,4}$, Clemens Mangler$^1$, Toma Susi$^1$, and Jani Kotakoski$^{1,*}$\\
$^1$University of Vienna, Faculty of Physics, Boltzmanngasse 5,\\ 1090 Vienna, Austria\\
$^2$University of Vienna, Vienna Doctoral School in Physics, Boltzmanngasse 5,\\ 1090 Vienna, Austria\\
$^3$School of Chemistry, Trinity College Dublin, College Green, Dublin 2, Ireland\\
$^4$CRANN and AMBER Research Centres, Trinity College Dublin, College Green,\\ Dublin 2, Ireland\\
$^*$Email: carsten.speckmann@univie.ac.at and jani.kotakoski@univie.ac.at}
\date{\today}


\begin{abstract}

    Electron irradiation-induced damage is often the limiting factor in imaging materials prone to ionization or electronic excitations due to inelastic electron scattering.
    Quantifying the related processes at the atomic scale has only become possible with the advent of aberration-corrected (scanning) transmission electron microscopes and two-dimensional materials that allow imaging each lattice atom.
    While it has been shown for graphene that pure knock-on damage arising from elastic scattering is sufficient to describe the observed damage, the situation is more complicated with two-dimensional semiconducting materials such as MoS$_2$.
    Here, we measure the displacement cross section for sulfur atoms in MoS$_2$ with primary beam energies between 55 and 90~keV, and correlate the results with existing measurements and theoretical models.
    Our experimental data suggests that the displacement process can occur from the ground state, or with single or multiple excitations, all caused by the same impinging electron.
    The results bring light to reports in the recent literature, and add necessary experimental data for a comprehensive description of electron irradiation damage in a two-dimensional semiconducting material.
    Specifically, the results agree with a combined inelastic and elastic damage mechanism at intermediate energies, in addition to a pure elastic mechanism that dominates above 80~keV.
    When the inelastic contribution is assumed to arise through impact ionization, the associated excitation lifetime is on the order of picoseconds, on par with expected excitation lifetimes in MoS$_2$, whereas it drops to some tens of femtoseconds when direct valence excitation is considered.

\end{abstract} 

\maketitle
\newpage

\section{Introduction}

    Although electron irradiation damage in (scanning) transmission electron microscopy [(S)TEM] has been investigated for decades~\cite{egerton_radiation_2004, egerton_control_2013}, its detailed quantification only became possible with two-dimensional (2D) materials~\cite{susi_quantifying_2019} that allow imaging all atoms in the structure with aberration-corrected~\cite{haider_spherical-aberration-corrected_1998,krivanek_towards_1999,kabius_first_2009} instruments.
    Beyond chemical etching~\cite{leuthner_scanning_2019} taking place due to non-ideal vacuum or sample contamination (which was recently shown not to be the case for MoS$_2$ even up to pressures of $10^{-7}$~mbar~\cite{ahlgren_atomic-scale_2022}), this damage is known to result from elastic and inelastic scattering of the imaging electrons from the material.
	During the elastic process, the electron scatters with a nucleus, transferring it a certain amount of kinetic energy, which causes a displacement if the material- and element-specific displacement energy threshold is exceeded.
	Such displacements dominate at higher electron energies, and have been studied for several 2D materials such as hexagonal boron nitride~\cite{meyer_selective_2009,jin_fabrication_2009,kotakoski_electron_2010}, graphene~\cite{meyer_accurate_2012,meyer_erratum_2013,susi_isotope_2016}, MoS$_2$~\cite{komsa_two-dimensional_2012,kretschmer_formation_2020} and MoSe$_2$~\cite{lehtinen_atomic_2015}.
	In contrast, inelastic scattering arises from the interaction of the beam electron with other degrees of freedom causing charging, ionization, electron excitations, or heating of the target, which can lead to weakening or breaking of bonds~\cite{egerton_control_2013}. 

    It has been well established that purely elastic knock-on describes electron beam damage in pristine graphene, and that lattice vibrations must be taken into account for an accurate description at electron energies close to the otherwise sharp damage onset determined by the displacement energy threshold of the material~\cite{meyer_accurate_2012,meyer_erratum_2013,susi_isotope_2016,chirita_three-dimensional_2022}.
	However, a similarly detailed description of the knock-on process is lacking for semiconducting or insulating 2D materials both due to the lack of experimental data and theoretical models.
    Nevertheless, it is clear that the simple elastic model that is sufficient for graphene can not comprehensively describe electron beam damage in an insulator such as hexagonal boron nitride~\cite{kotakoski_electron_2010}.
    Until now, the role of the inelastic scattering has been approached indirectly for the semiconducting transition metal dichalcogenides (TMDs) MoS$_2$~\cite{algara-siller_pristine_2013} and MoSe$_2$~\cite{lehnert_electron_2017} by comparing damage with and without a protective graphene layer.
    Meanwhile, the ground state displacement threshold energies (corresponding to the elastic case) for several TMDs have also been estimated through first-principles atomistic simulations~\cite{komsa_two-dimensional_2012,yoshimura_first-principles_2018}.
    Only recently, the joint contribution of elastic and inelastic scattering to the damage in MoS$_2$ has been directly considered by Kretschmer et al. combining experiments and simulations~\cite{kretschmer_formation_2020}. 
    Yoshimura et al. also recently developed a model combining density-functional theory (DFT) calculations and quantum electrodynamics (QED) to describe damage in non-conductive 2D materials~\cite{yoshimura_quantum_2022}.
    However, the experimentally studied primary beam energy range has been limited to $\leq 80$~keV, precluding sulfur displacements from the electronic ground state, and no satisfactory agreement has been found between experimental results and a model with a direct physical interpretation.
	
    In this work, we measure the displacement cross section in monolayer MoS$_2$ at acceleration voltages of $55-90$~kV using aberration-corrected STEM imaging combined with a convolutional neural network-assisted analysis.
	Our results show a clear increase for the cross section values above 80~kV, which allows a direct comparison to theory including displacements also from the electronic ground state.
	We show that describing the inelastic contribution as impact ionization of the sulfur atom leads to a satisfactory agreement with the experimental data and a deexcitation time of up to picoseconds, in agreement with literature values~\cite{korn_low-temperature_2011,lin_physical_2017,palummo_exciton_2015}.
	In contrast, describing it via valence excitation based on DFT calculations leads to a better agreement, but only with deexcitation lifetimes as short as some tens of femtoseconds, and only if the deexcitation is described as a collective process.
	Overall, the results presented here provide the first experimental data that allows the development of a comprehensive model for describing electron irradiation damage in a 2D semiconducting material, while also giving the first reliable indications of the relevant underlying physical phenomena.

\section{Results and Discussion}

\begin{figure*}[ht!]
\centering
\includegraphics[width=0.8\textwidth]{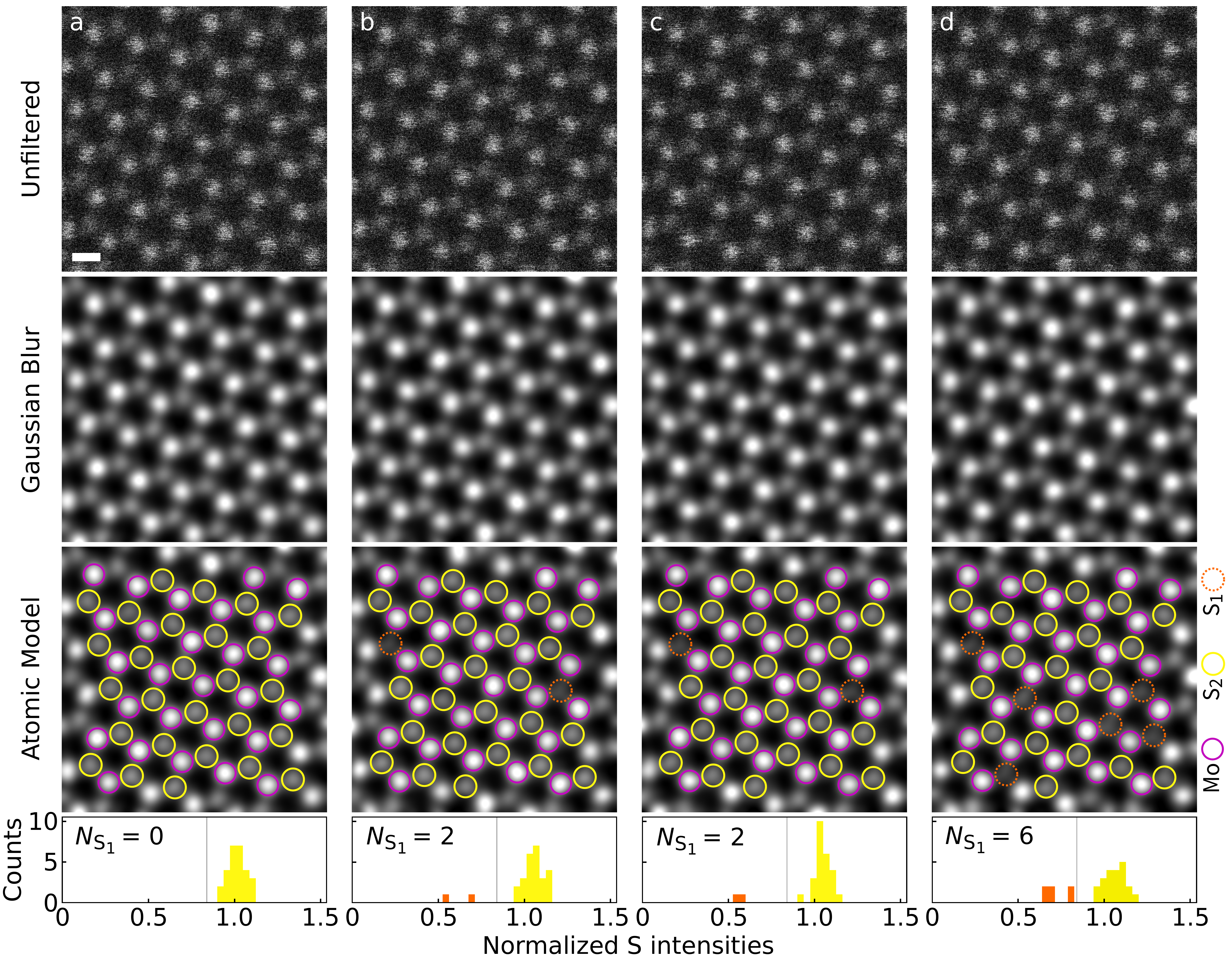}
    \caption{{\bf STEM-HAADF image series of MoS$_2$.} 
    Four subsequent image frames (a-d) recorded with a dwell time of 3~$\mu$s/px, a flyback time of 120~$\mu$s and $512~\mathrm{px}\times 512~\mathrm{px}$ resulting in a frame time of 0.85~s with an additional 10~ms between the frames. Images in the top row are as-recorded, whereas a Gaussian blur (9~px) has been applied on the images in the second row, with an additional overlay indicating the positions of the atomic sites in the third row.
    Solid purple circles indicate molybdenum atoms while solid yellow and dotted orange circles highlight a column of two sulfur atoms and a single sulfur atom/single sulfur vacancy, respectively.
    The histograms show the corresponding sulfur site intensities (marked in the same color as their respective positions in the third row), normalized to the mean of each frame individually.
    Additionally, the number of single sulfur vacancies $N_{\mathrm{S_1}}$ in each image is written into the histogram.
    The scale bar is 2~{\AA}.}
\label{fig:Imageseries}
\end{figure*}

    Monolayer MoS$_2$ samples were grown by chemical vapor deposition (CVD), and transferred onto a TEM grid (Au grid with a Quantifoil carbon film) for STEM imaging (see Methods for details).
    Atomic resolution image series were recorded for acceleration voltages between 55 and 90~kV with a step size of 5~kV (ca. 100 series per voltage) to allow observing the displacement of sulfur atoms as a function of the electron dose.
	A typical recorded image series is shown in Fig.~\ref{fig:Imageseries}.
    The first row in this figure shows four consecutive unfiltered images of atomically resolved MoS$_2$, where the lattice sites can be easily identified due to the $Z$-contrast~\cite{pennycook_z-contrast_1989} of the STEM high angle annular dark field (HAADF) imaging mode (the contrast is approximately proportional to the atomic number $Z^{1.64}$~\cite{krivanek_atom-by-atom_2010}).
    A Gaussian blur was applied to the images in the second row, and in the third row those atomic positions that were taken into consideration during the following analysis are marked with circles.
    The first image (Fig.~\ref{fig:Imageseries}a) shows a completely intact hexagonal lattice structure of Mo atoms alternating with S sites (each having two S atoms on top of each other), indicated with purple and yellow circles, respectively.
    In the subsequent images, some of the sulfur sites exhibit reduced intensity (dotted orange circles), suggesting a missing sulfur atom at this position.
    Assuming that the number of any newly created vacancies is entirely stochastic and not dependent on the environment, a linear dependency is expected for the average number of new defects as a function of electron dose~\cite{meyer_accurate_2012,meyer_erratum_2013}.
    We note that since all sulfur sites in the structure are equivalent, the field of view used in the measurement has no statistical influence on the uncertainty of the experimental results.
    However, this is only true as long as the local vacancy concentration remains low.
    For this reason, each experiment (image sequence) was ended when more than two sulfur vacancies appeared adjacent to one another as next nearest neighbors (as in Fig.~\ref{fig:Imageseries}d).
	Additionally, to avoid influence from contamination on the results, all image series with contamination were excluded from the analysis, as was also the area in each image closer to the edge than the projected Mo-S site distance. 
    The histograms (last row in Fig.~\ref{fig:Imageseries}) show the sulfur site intensities of the corresponding image, that allow us to count the number of sulfur vacancies.
    
    To assist defect recognition, a convolutional neural network (CNN) was used to identify the atomic positions for which the intensities are calculated.
    As indicated by the different colors, all intensities located below a certain threshold (shown with the grey line) are counted as vacancies.
	Due to variations in the intensity from one image series to the next, this threshold was set manually for each series.
    Despite the help of the CNN, this process is still estimated to lead to the largest uncertainty in the analysis, which was assessed by having the same data analyzed by two independant people.
	For the analysis, the number of defects present in the first frame is used as a reference, and only the change of the number of S vacancies during the experiment is used.
    Note that this number can also be negative when the identification of a site changes during the analysis due to changing intensity (either due to vacancy healing or due to inaccuracy in estimating the intensity due to noise).
	As the result, each image series yields an individual ratio for the number of created sulfur vacancies per image, obtained by fitting a linear regression to the data.
	Details on the CNN and the analysis method are described in the Methods section.
	
\begin{figure}[hb!]
\centering
\includegraphics[width=0.6\textwidth]{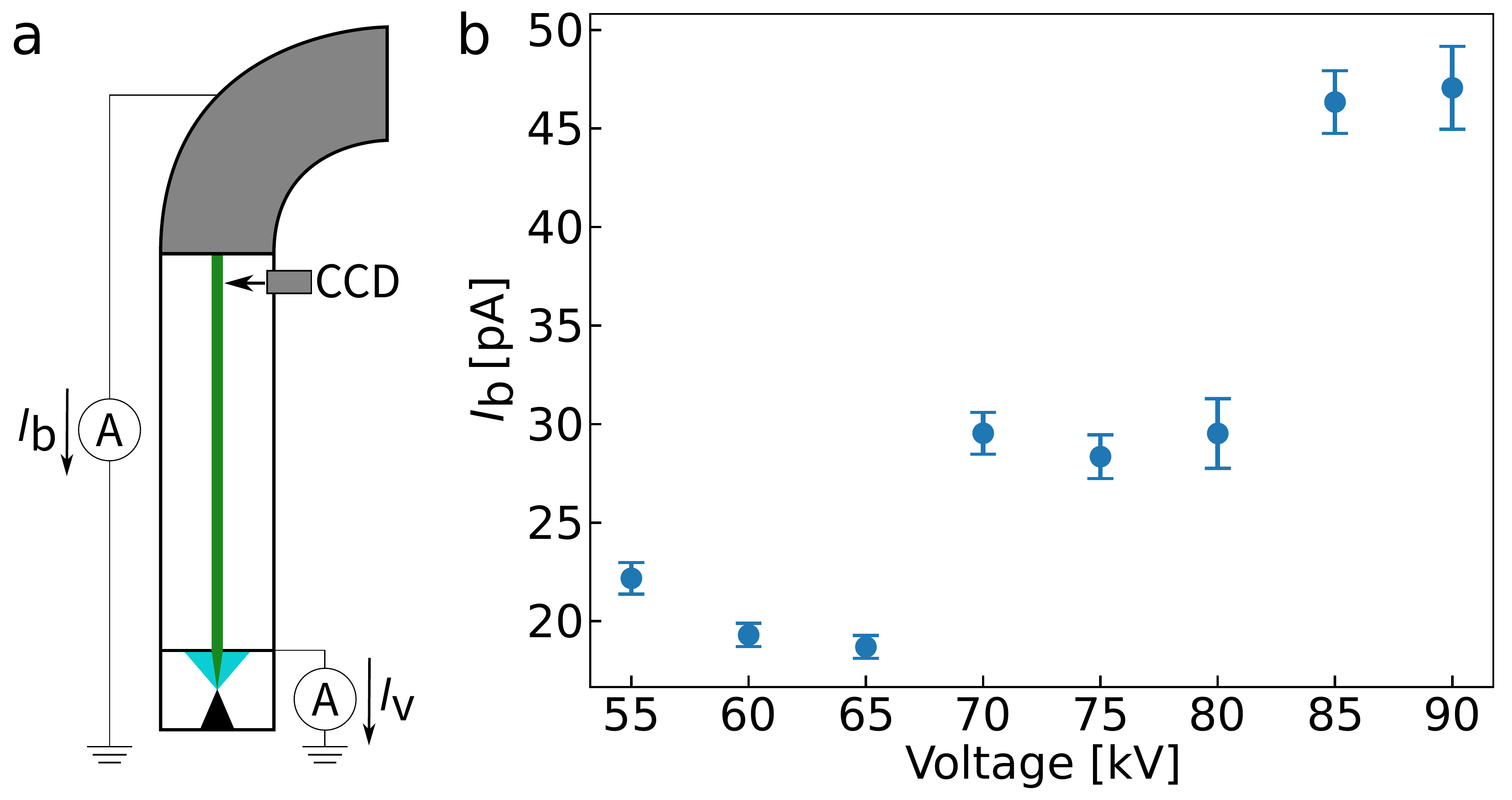}
    \caption{{\bf Beam current measurements.} 
    (a) Simplified schematic illustration of the measurement system. 
    The electron source is shown at the bottom in black, with those emitted electrons blocked by the VOA shown in cyan and those passing through the column shown in green.
    The beam current $I_{\mathrm{b}}$ is measured using the drift tube of the electron energy-loss spectrometer (at the top), while the VOA current $I_{\mathrm{v}}$ is directly measured simultaneously.
    Dark current subtracted from $I_{\mathrm{b}}$ was measured by inserting the Ronchigram charge coupled device (CCD) camera to block the beam.
    (b) Mean values for $I_{\mathrm{b}}$ as a function of the electron acceleration voltage.}
\label{fig:Beamcurrent}
\end{figure}

    For each recorded image, we also store in the metadata the simultaneously measured current ($I_{\mathrm{v}}$) at the virtual objective aperture (VOA) of the microscope.
    This corresponds to the part of the electron probe that is cut off by the aperture and never encounters the sample.
    To estimate the actual beam current $I_{\mathrm{b}}$, which is needed to calculate the electron dose per image, we carried out a set of calibration measurements, as schematically illustrated in Fig.~\ref{fig:Beamcurrent}a.
    In these measurements, instead of using a Faraday cup, as is typical with other instruments, $I_{\mathrm{b}}$ was recorded as the current between the microscope ground and the drift tube for the electron energy loss spectrometer (without a sample), while simultaneously measuring $I_{\mathrm{v}}$.
    The dark current that was subtracted from $I_{\mathrm{b}}$ was recorded by blocking the electron beam with the Ronchigram camera.
    Emission of the electron gun was altered by changing the extraction voltage during the measurement to establish the linear dependency of $I_{\mathrm{b}}$ on $I_{\mathrm{v}}$ to allow estimating the actual electron dose for each image.
    Note that the decay of the beam current of a cold field emission gun due to the absorption of gas molecules is known~\cite{williams_transmission_2009}, but has no influence on our results, because the beam current is measured separately for each recorded image.
    We point out that it is typical to measure the beam current only sporadically and for short times, which based on our experiments can lead to significant errors in the estimated beam current.
    Therefore, we measure the beam current separately for each microscope alignment, and do the measurement for long enough (ca. 40 measurements and a total of up to several hours for each alignment) to ensure that its variation is correctly captured.
    The results from the beam current measurements are shown in Fig.~\ref{fig:Beamcurrent}b.
    
\begin{figure}[hb]
\centering
\includegraphics[width=0.6\textwidth]{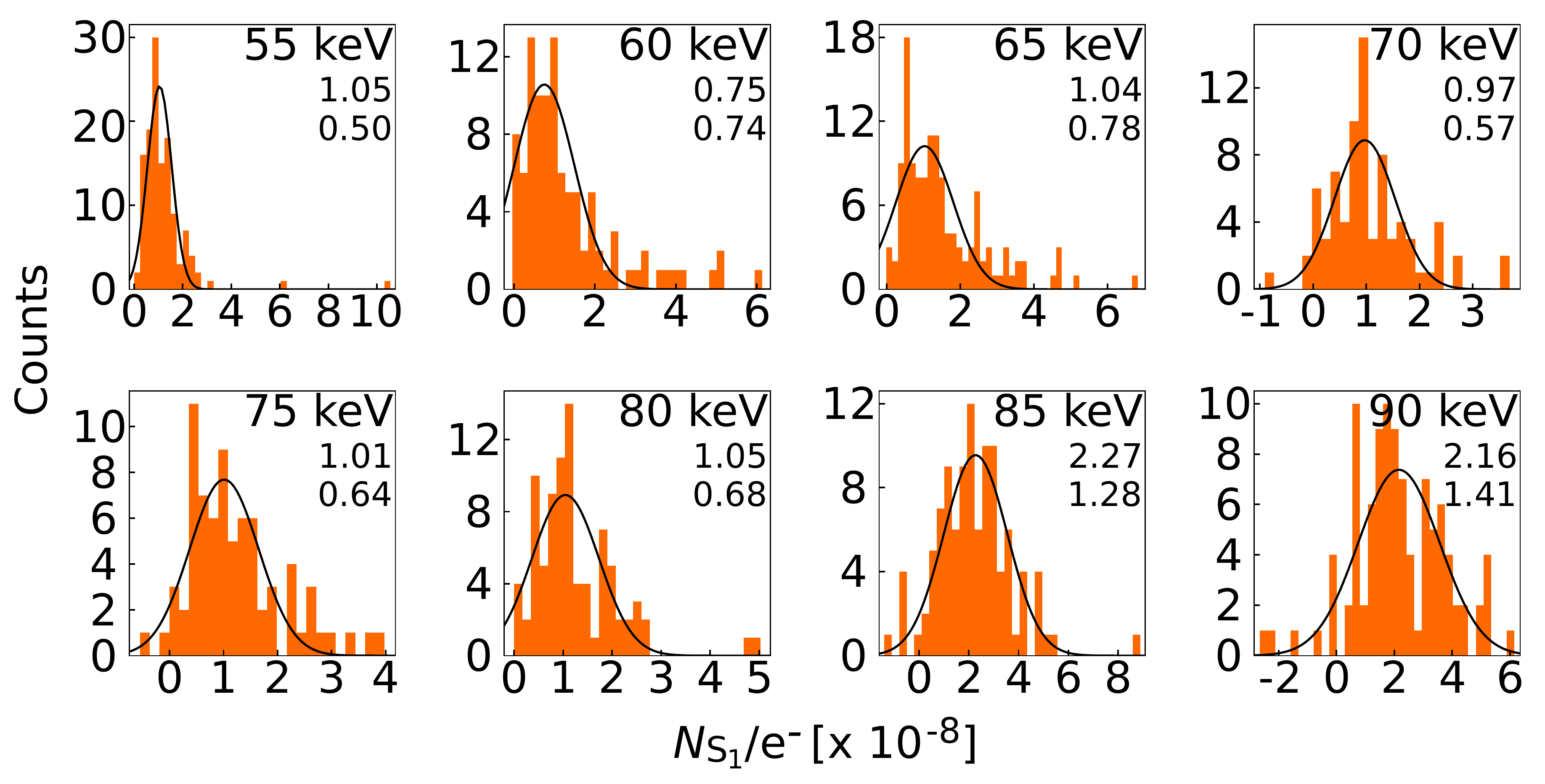}
\caption{{\bf Statistics of sulfur vacancy formation.}
    Histograms of the average number of single sulfur vacancies created per impinging electron for the energies stated in the top right corner of each plot.
    Here, one count represents the results of a single image series.
    The solid lines correspond to normal distributions with means $\bar{x}$ and standard deviations $\Delta x$ given in the top-right corner of each plot underneath the electron energies.}
\label{fig:Datasets}
\end{figure}

    Combining the number of created vacancies from the analysis of the image series with the measured $I_{\mathrm{v}}$ currents and the ratio $I_{\mathrm{b}}/I_{\mathrm{v}}$ allow us to calculate the average number of vacancies created per impinging electron $\left( N_{\mathrm{S_1}}/\mathrm{e}^- \right)$ for each image series.
    These data are summarised in histograms in Fig.~\ref{fig:Datasets}.
    Although the knock-on process of a single atom is Poisson distributed over electron dose~\cite{susi_siliconcarbon_2014}, we measure here the average from several such processes.
    According to the central limit theorem~\cite{polya_uber_1920}, the measured expectation values collectively follow the normal distribution, as seen in Fig.~\ref{fig:Datasets}.
    This allows us to calculate the knock-on cross section as $\sigma_\mathrm{tot} = \bar{x}\rho$, where $\bar{x}$ is the mean of the normal distribution for $\left( N_{\mathrm{S_1}}/\mathrm{e}^- \right)$ and $\rho$ is the areal density of the sulfur site in MoS$_2$ (i.e., the inverse of the unit cell area calculated with a lattice constant of $3.19$~{\AA}~\cite{ahmad_comparative_2014}).

\begin{figure}[ht]
\centering
\includegraphics[width=0.6\textwidth]{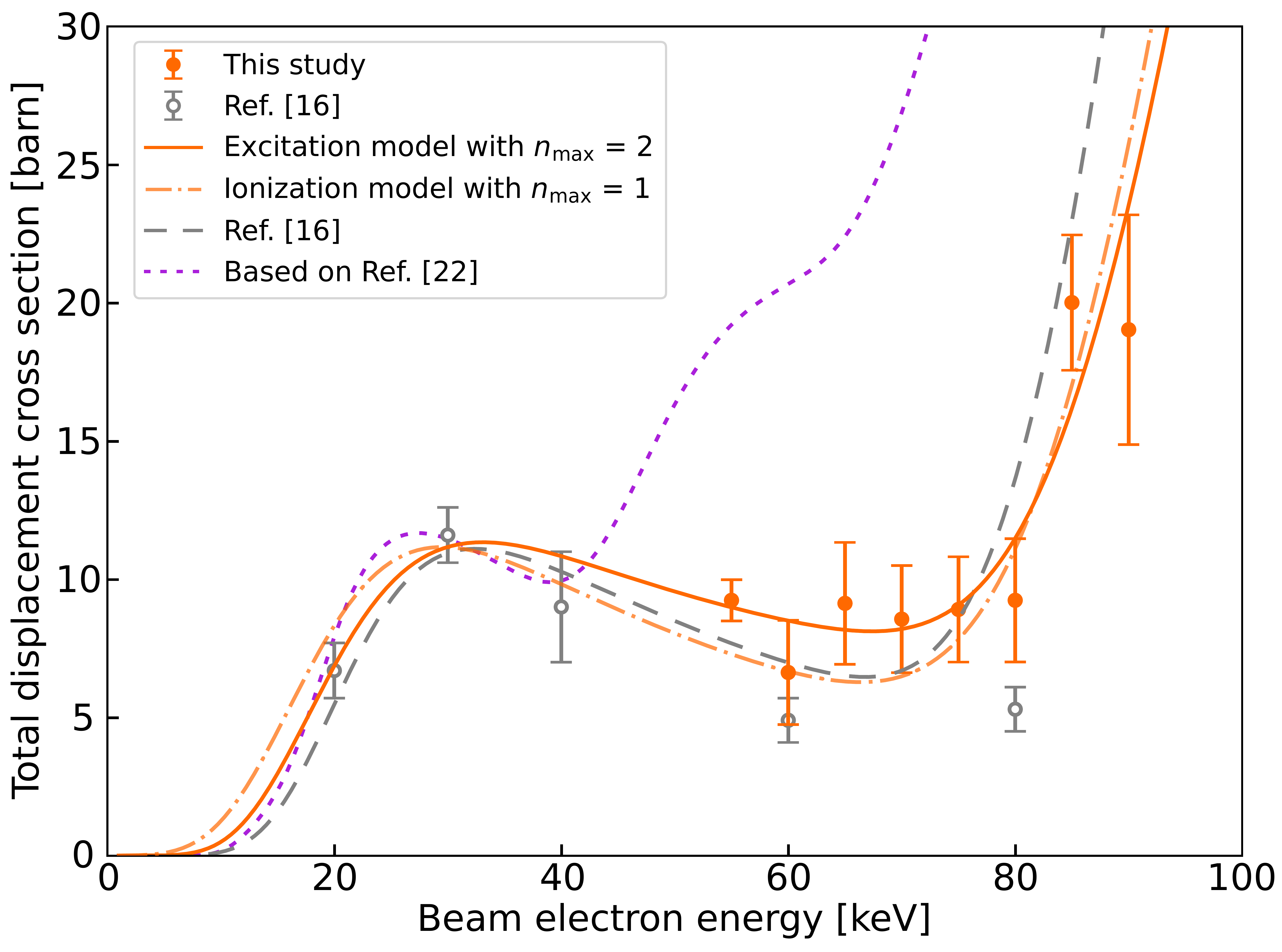}
\caption{{\bf Experimental displacement cross section values compared to the models.}
    Total displacement cross section of single sulfur atoms in MoS$_2$ as a function of electron energy measured during the course of this work (filled orange circles) and from Ref.~\cite{kretschmer_formation_2020} (open grey circles).
    The dashed grey line shows the model fit from Ref.~\cite{kretschmer_formation_2020}, whereas the dotted purple line corresponds to the model from Ref.~\cite{yoshimura_quantum_2022} taking into account all excited states included in the model.
    Dash-dotted orange line corresponds to our model assuming impact ionization based on the Bethe cross section and the solid orange line corresponds to our model assuming electronic excitations up to two excited states.
    Parameters for our two models with variable numbers of excited electrons are listed in Table \ref{tbl:fits}.}
\label{fig:Fits}
\end{figure}

    As was already pointed out, although for graphene elastic scattering is sufficient to describe the observed knock-on damage, this does not hold for MoS$_2$.
    This is immediately apparent from the experimental data (Fig.~\ref{fig:Fits}), since for a purely elastic description, the cross section values should gradually increase almost exponentially within the acceleration voltage range studied here.
    In contrast, the results for MoS$_2$ show a nearly constant value in the range of $55-80$~kV, and only start to increase above 80~kV.
	The purely elastic knock-on cross section $\sigma_{\mathrm{KO}}$ can be described by the formalism of McKinley and Feshbach~\cite{mckinley_coulomb_1948} modified to account for a target atom vibrating in the out-of-plane direction~\cite{zobelli_electron_2007}, where
\begin{widetext}
\begin{equation}
\begin{aligned}
\sigma_{\mathrm{KO}} (T, E_\mathrm{d}) = 4 \pi \left( \frac{Z e^2}{4 \pi \epsilon_0 2 m_0 c^2} \right)^2 \frac{1 - \beta (T)^2}{\beta (T)^4} \Biggl\{ \frac{E_\mathrm{max} (T,v)}{E_\mathrm{d}} - 1 - \beta (T)^2 \ln{ \left( \frac{E_\mathrm{max} (T,v)}{E_\mathrm{d}} \right)} \\ + \pi \frac{Z e^2}{\hbar c} \beta (T) \left[ 2 \sqrt{ \frac{E_\mathrm{max} (T,v)}{E_\mathrm{d}} } - \ln{ \left( \frac{E_\mathrm{max} (T,v)}{E_\mathrm{d}} \right)} - 2 \right] \Biggr\},
\end{aligned}
\label{eq:MF}
\end{equation}
\end{widetext}
    with $E_\mathrm{d}$ being the displacement threshold energy (the lower boundary of energy that needs to be transferred to the atomic nucleus with mass $M$ and atomic number $Z$ to cause a knock-on event).
	$E_\mathrm{max}$ is the maximum kinetic energy that an electron with kinetic energy $T$, elementary charge $e$ and electron rest mass $m_0$ can transfer to the nucleus moving at a velocity $v$, $\epsilon_0$ is the vacuum permittivity, $c$ the speed of light in vacuum, $\hbar$ the reduced Planck constant and $\beta (T) = \sqrt{1 - (1 + T/(m_0c^2))^{-2}}$ is the velocity of the electron in units of $c$.
	The velocity of the nucleus is determined via the material's phonon dispersion, affecting $E_\mathrm{max}$ and therefore $\sigma_{\mathrm{KO}}$ as described in~\cite{susi_isotope_2016}.
	The phonon dispersion of the material was calculated as described in the Methods.
	
	Inelastic contributions to the process can be included via a probability $P_n$, which can be constructed as the product of the probability of one electron impinging on the sample exciting $n$ electrons and the probability of these excitations to live long enough to contribute to the displacement process.
	Here, we discuss two different ways to take this into account.
    In the first one, where we write $P_n = P^\mathrm{I}_n$, we follow Kretschmer et al.~\cite{kretschmer_formation_2020}, who assumed that the excitation can be described based on the ionization cross section according to Bethe~\cite{bethe_zur_1930}.
    In the second one ($P_n = P^\mathrm{E}_n$) we follow Yoshimura et al.~\cite{yoshimura_quantum_2022} who calculated excitation probabilities directly from first principles.
    In both cases, we assume that the deexcitation takes place collectively (the system returns to ground state with a single deexcitation event), and follows Poisson statistics where the defining parameter is the excitation lifetime $\tau_n$ that leads to an exponentially decreasing probability for the excitation to exist long enough for the displacement to take place.
    In contrast, Kretschmer et al.~\cite{kretschmer_formation_2020} replaced the deexcitation probability with an arbitrary efficiency factor, whereas Yoshimura et al.~\cite{yoshimura_quantum_2022} considered the deexcitation of each of the $n$ excitations independently.
    The benefit of our approach over the one by Kretschmer et al. is that in our case all parameters of the model are physically motivated.
    In comparison, as will be discussed below, the assumption of independent deexcitation in the model proposed by Yoshimura et al. leads to a stark disagreement with our experimental data.

    For completeness, the probabilities for an electronic excitation to occur and to exist long enough to influence the knock-on process as a function of the kinetic energy of the impinging electron $T$ become now
\begin{equation}
    \begin{array}{ll} P^\mathrm{I}_{n > 0}(T) = \sigma_{\mathrm{ve}} (T,n) \rho \exp \left( \frac{-t_{\mathrm{disp}}}{\tau_{n}} \right) \quad \text{and} \\
         P^\mathrm{E}_{n > 0}(T) =  \frac{S^{n}}{n!} \exp \left( -S \right) \exp \left( \frac{-t_{\mathrm{disp}}}{\tau_{n}} \right) \end{array}
\end{equation}
for $n>0$ excited states. Probability for the system to be in ground state is
\begin{equation}
	 P_{n=0}(T) = 1 - \sum_{i=1}^{n_{\mathrm{max}}} P_{n=i}(T).
\label{eq:Probabilities}
\end{equation}
    $t_{\mathrm{disp}}$ is the time the atom needs to be displaced assuming that the impinging electron transferred an energy of exactly $E_\mathrm{d}$ (we additionally assume, following Ref.~\cite{kretschmer_formation_2020} the atom to be displaced when it has moved 4.5~{\AA} from its original position).
	$\sigma_{\mathrm{ve}}$ is the relativistic Bethe inelastic scattering cross section in a material with $n_{\mathrm{ve}}$ valence electrons as described in Ref.~\cite{bui_electron_2023}

\begin{widetext}
\begin{equation}
\begin{aligned}
	\sigma_{\mathrm{ve}} (T) = \frac{\pi e^4 b_{\mathrm{ve}} n_{\mathrm{ve}}}{\left(4 \pi \epsilon_0\right)^2 T E_{\mathrm{c}}} \left[\ln{ \left(\frac{c_{\mathrm{ve}} T}{E_{\mathrm{c}}} \right)}\right.\\
	\left.- \ln{\left(1-\beta (T)^2 \right)} - \beta (T)^2 \right],
\end{aligned}
\label{eq:Bethe}
\end{equation}
\end{widetext}
    where $E_{\mathrm{c}}$ is the ionization energy, and $b_{\mathrm{ve}}$ and $c_{\mathrm{ve}}$ are constants relevant to the valence shell that we obtain by fitting the sum of the first and the second ionization cross section to the electron impact ionization data of sulfur~\cite{tawara_total_1987}.
    The resulting fit parameters are $b_{\mathrm{ve}} = 0.381 \pm 0.002$ and $c_{\mathrm{ve}} = 0.938 \pm 0.004$ with ionization energies of $E_{\mathrm{c}1} = 10.36$~eV and $E_{\mathrm{c}2} = 23.33$~eV~\cite{kelly_atomic_1982}.
    $S$ is the sum of all probabilities for all possible transitions from the valence into the conduction band, as described in Ref.~\cite{yoshimura_quantum_2022}.

	The total displacement cross section can then be written as
	
\begin{widetext}
\begin{equation}
\begin{aligned}
\sigma_{\mathrm{tot}} (T, E_\mathrm{d}) = P_{n=0}(T) \sigma_{\mathrm{KO},n=0}(T, E_{\mathrm{d},n=0}) \\
	+ \sum_{i=1}^{n_{\mathrm{max}}} P_{n=i}(T) \sigma_{\mathrm{KO},n=i}(T, E_{\mathrm{d},n=i}), 
\end{aligned}
\label{eq:sigma_tot}
\end{equation}
\end{widetext}
	which was used to fit our experimental data. 
	For fitting the models (the ionization model using $P_n^\mathrm{I}$ and the excitation model using $P_n^\mathrm{E}$), we also include the cross section values below 55~kV from Ref.~\cite{kretschmer_formation_2020} since our data is limited to a voltage range of $55-90$~kV.
	However, we double the uncertainties given by Kretschmer et al. since they only reported uncertainties arising from the statistical sample size.
    Results of the fits are shown in Fig.~\ref{fig:Fits}.	

\begin{table*}[h!]
\caption{{\bf Displacement thresholds and excitation lifetimes.}
    Parameters obtained by fitting the theoretical models to the experimental data as shown in Fig.~\ref{fig:Fits} together with the corresponding values from the literature.}
\vspace{.4cm}
\begin{tabular}{|c|c|c|c|c|c|c|c|}\hline
  & $E_{\mathrm{d,gs}}$ (eV) & $E_{\mathrm{d},1}$ (eV) & $E_{\mathrm{d},2}$ (eV) & $E_{\mathrm{d},3}$ (eV) & $\tau_{1}$ (fs) & $\tau_{2}$ (fs) & $\tau_{3}$ (fs) \\\hline\hline
    Ref.~\cite{kretschmer_formation_2020} & 6.5$^\mathrm{t}$ & 4.8$^\mathrm{t}$; 1.5$^\mathrm{f}$ & 3.5$^\mathrm{t}$ & & & & \\\hline
 Ref.~\cite{yoshimura_quantum_2022} & 6.92$^\mathrm{t}$ & 5.04$^\mathrm{t}$ & 3.16$^\mathrm{t}$ & 1.28$^\mathrm{t}$ & 81$^\mathrm{f}$ & 81$^\mathrm{f}$ & 81$^\mathrm{f}$ \\\hline
    Ionization $n_{\mathrm{max}} = 1$ & 6.8 $\pm$ 0.1 & 1.4 $\pm$ 0.1 & & & $\gtrsim 5 \times 10^{3}$ & & \\\hline
    Excitation $n_{\mathrm{max}} = 1$ & 7.0 $\pm$ 0.1 & 0.9 $\pm$ 0.1 & & & 37 $\pm$ 1 & & \\\hline
    Excitation $n_{\mathrm{max}} = 2$ & 6.9 $\pm$ 0.2 & 3.7 $\pm$ 3.6 & 1.3 $\pm$ 0.1 &  & 20 $\pm$ 9 & 38 $\pm$ 1 &  \\\hline
\end{tabular}
\\
    $^\mathrm{t}$ Theoretically estimated value\\
    $^\mathrm{f}$ Value obtained from a fit
\label{tbl:fits}
\end{table*}

    As can be seen from Fig.~\ref{fig:Fits}, both our models used here reasonably agree with the experimental data.
    As expected since they use the same inelastic scattering cross section model, the difference between the ionization model and the model by Kretschmer et al.~\cite{kretschmer_formation_2020} is the smallest.
    For the displacement threshold from the ground state $E_{\mathrm{d,gs}}$ we obtain values between $6.8-7.0$~eV with all models, similar to the DFT values reported in Refs.~\cite{komsa_two-dimensional_2012,yoshimura_first-principles_2018,yoshimura_quantum_2022} and somewhat above the value of Kretschmer et al.~\cite{kretschmer_formation_2020}.
    In the case of the ionization model, the displacement threshold from the excited state (with ionized S) is $1.4 \pm 0.1$~eV, close to the value of Ref.~\cite{kretschmer_formation_2020}, similarly obtained through a fit and mostly determined by their experimental data at the lowest acceleration voltages.
    The lifetime for the ionization resulting from this model is $\gtrsim 5$~ps, which is similar to what was reported for excitons in MoS$_2$~\cite{korn_low-temperature_2011,lin_physical_2017,palummo_exciton_2015}.
    We also point out that defects have been shown to extend the lifetime of excitons in TMDs by up to an order of magnitude~\cite{liu_neutral_2019}.
    Adding multiple ionized states ($n>1$) did not result in an improved fit.
    
    For the excitation model, the experimental data can be fitted with either one or two excited states, but we only show the plot for the latter case due to its better fit to the experimental data.
    With this model, the excited state displacement threshold energies become $3.7\pm 3.6$~eV and $1.3\pm 0.1$~eV, for single and double excitations, respectively.
    The very large uncertainty for the single excited state results from the relatively constant displacement cross section at intermediate energies, which allows fitting the data with different threshold energies varying the corresponding lifetime. 
    The lifetimes for these excited states become $20\pm 9$~fs and $38~\pm 1$~fs, which are the same order of magnitude as the 81~fs obtained in Ref.~\cite{yoshimura_quantum_2022}, where all excitations were assumed independent and only $n\geq 3$ was fitted to the experimental results of Ref.~\cite{kretschmer_formation_2020}.
    It is worth noting that this time scale is longer than the lifetime of a core hole, but much shorter than the lifetimes of any other expected excitation.
    In essence, a lifetime that is too brief means that the used description of inelastic scattering gives a probability that is too high.
    Comparing these results suggests that the true inelastic contribution to the electron-beam damage cross section in MoS$_2$ is similar to the probability for direct ionization, but smaller than the probability for valence excitation. 

\section{Conclusions}

	In conclusion, our results confirm that inelastic scattering needs to be taken into account to accurately describe the knock-on process of S atoms from MoS$_2$ under electron irradiation.
    Building on recent prior works~\cite{kretschmer_formation_2020,yoshimura_quantum_2022}, we compare the experimental measurements to theoretical models, and show that the data can be qualitatively described by either assuming electron impact ionization of the sulfur atom or by assuming valence excitation of the material.
    The data is consistent with a sum of displacement cross sections for MoS$_2$ at the ground state and one or more excited states with the lowest displacement threshold energy being ca. 1.3~eV (highest relevant excitation) and the highest, corresponding to the ground state, being ca. 6.9~eV.
    However, only with the impact ionization model the resulting lifetime becomes similar to excitation lifetimes reported in the literature (up to tens of picoseconds)~\cite{korn_low-temperature_2011,lin_physical_2017,palummo_exciton_2015}, whereas the valence excitation leads to unexpectedly short lifetimes (tens of femtoseconds).
    Overall, the models presented here describe the experimentally measured displacement cross section in a 2D semiconducting material with only physically motivated parameters, which allows direct interpretation of the underlying physics opening the door to further theoretical work.
    Nevertheless, additional experimental data especially at lower electron energies are necessary to confirm that the cross section indeed increases gradually as suggested by the combination of inelastic and elastic scattering discussed here, rather than arising for example from chemical changes in the material under electronic excitation.

\section*{Methods}

        {\bf Sample preparation} The MoS$_2$ sample was grown on SiO$_2$ via chemical vapor deposition (CVD)~\cite{obrien_transition_2014}, and was afterwards transferred in air onto a gold transmission electron microscopy grid with a holey membrane of amorphous carbon (Quantifoil R 1.2/1.3 Au grid) with a method similar to that described in Ref.~\cite{meyer_hydrocarbon_2008}.
        The sample was introduced to vacuum and baked overnight at ca. 150$^\circ$C before measurements were conducted.
        In between the measurements, the sample was stored in the CANVAS ultra-high vacuum system at the University of Vienna~\cite{mangler_materials_2022}.

        {\bf Scanning transmission electron microscopy} Measurements were carried out with a Nion UltraSTEM 100, an aberration-corrected scanning transmission electron microscope with electron acceleration voltages in the range from 55 to 90~kV and a probe size of $\sim$1~\r{A}.
        The instrument is equipped with a cold field emission gun, and images were recorded using high-angle annular dark field (HAADF) and medium-angle annular dark field (MAADF) detectors (for 60~kV only) with collection angles of $80-300$~mrad and $60-200$~mrad, respectively.
        The base pressure at the sample was below 10$^{-9}$~mbar at all times.
        Imaging parameters were chosen to minimize the dose per frame while retaining enough signal to reliable recognize the atomic sites (dwell time of 3~$\mu$s/px, flyback time of 120~$\mu$s and $512~\mathrm{px}~\times~512~\mathrm{px}$ images) resulting in a total frame time of 0.85~s with an additional 10~ms between the frames.
        The field of view for each image was 1.9~nm and the beam convergence angle was 30~mrad.
        Image series acquisition was stopped as soon as a small vacancy cluster (more than two missing S atoms at next-nearest neighboring lattice sites) was observed.
        
        {\bf Phonon density of states calculation} The phonon calculations were performed using the density functional perturbation theory (DFPT)~\cite{baroni_phonons_2001,gonze_dynamical_1997} approach implemented in the ABINIT code~\cite{gonze_abinit_2009}.
        First the lattice structure was optimized for both cell size and atomic positions up to an energy difference of $1.0\times 10^{-10}$~Ha.
        We used a $\mathbf{k}$-point mesh of $20\times 20\times 1$ with an energy cut-off of 55~Ha.
        The local density approximation and the Troullier–Martins norm conserving pseudo-potentials~\cite{troullier_efficient_1991} were used to describe
the exchange correlation potential.
        For calculating the Hessian and dynamic matrix, the ground state wave functions were converged to $10^{-18}$~Ha$^2$ with a $\mathbf{k}$-point mesh of $40\times 40\times 1$.
        Density of states was interpolated using the Gaussian method with a smearing value of $6.5\times 10^{-5}$~Ha.

	{\bf Data analysis} The detection of sulfur vacancies was assisted by a CNN similar to the one described in Ref.~\cite{trentino_atomic-level_2021}.
	The CNN calculates a heat map of probabilities of an atom being present at each pixel in each image, and converts this map into a set of atomic positions by locating peaks in the heat map.
	Later, a model image is created by overlapping the calculated atomic positions with the actual image and modelling the shape of the electron probe as a superposition of Gaussians.
The optimization of the model is achieved by minimizing the intensity difference to the recorded image as described in Ref.~\cite{postl_indirect_2022}.
	These intensities were then used to differentiate between the two different lattice sites, and to identify vacancies by setting an intensity threshold, labeling all sulfur lattice positions with a lowered intensity as vacancies.
	Image series containing any contamination or other disorder were discarded from the analysis.
	Similarly, only atoms at least one projected bond length away from the edge of the image were taken into account to rule out the influence of possible contamination, adatoms, and vacancy clusters located directly outside the field of view.
	The highest uncertainty of this analysis, apart from the subjective interpretation of the researcher, arose from the Gaussian fit in Fig.~\ref{fig:Datasets} (up to 10~\%).
	To account for the uncertainty caused by the subjective interpretation of the researcher, the analysis for the datasets was done by two researchers independently.
	The total error was estimated by multiplying the uncertainty caused by the Gaussian fit with a factor of $2.5$ to contain both sources of uncertainty.
	All other uncertainties were deemed negligible in comparison to these errors.

\section*{Acknowledgments}

We acknowledge funding from the Austrian Science Fund (FWF) through doctoral college Advanced Functional Materials--Hierarchical Design of Hybrid Systems (DOC~85), from the Vienna Doctoral School in Physics and from the European Research Council (ERC) under the European Union's Horizon 2020 research and innovation programme Grant agreement No. 756277-ATMEN.

\bibliographystyle{elsarticle-num}
\bibliography{references}

\end{document}